\documentclass[global,twocolumn]{svjour}

\usepackage{graphicx,amsmath}
\usepackage{hyperref}

\journalname{arXiv}

\begin{document}

\title{Two-color photoionization of calcium using SHG and LED light}

\author{Carsten Schuck\inst{1}\fnmsep\thanks{Fax: +34-93-553-4000, E-mail: carsten.schuck@icfo.es} \and Felix Rohde\inst{1} \and Marc Almendros\inst{1} \and Markus Hennrich\inst{1,2} \and
J\"{u}rgen Eschner\inst{1}}

\institute{ICFO-Institut de Ci\`{e}ncies Fot\`{o}niques, E-08860
Castelldefels, Barcelona, Spain. \and Institut fu¨r
Experimentalphysik, Universita¨t Innsbruck, Technikerstrasse 25,
A-6020 Innsbruck, Austria}
\date{\today}

\maketitle

\begin{abstract} We present a photoionization method to load single
$^{40}$Ca ions in a linear Paul trap from an atomic beam. Neutral
Ca I atoms are resonantly excited from the ground state to the
intermediate 4s4p $^1$P$_1$-level using coherent 423nm radiation
produced by single-pass second harmonic generation in a
periodically poled KTiOPO$_4$ crystal pumped with an 120mW
extended cavity diode laser. Ionization is then attained with a
high-power light emitting diode imaged to the trap center, using
an appropriately designed optical system composed of standard
achromatic doublet lenses. The setup simplifies previous
implementations at similar efficiency, and it hardly requires any
maintenance at all.\vspace{2mm}
\end{abstract}

\section{Introduction}

Trapped ions are considered to be one of the most attractive
systems for the implementation of scalable quantum information
processing \cite{hb08}. In particular $^{40}$Ca$^+$ has shown to
be a well suited qubit candidate because entangled states of two
or more qubits can be prepared, coherently controlled and read out
with high fidelity \cite{bw08}. For practical realizations,
calcium ions have the important advantage that all light sources
required for cooling and information processing are available as
(frequency doubled) diode lasers.

\begin{figure}[!t]
\begin{center}
\includegraphics[width=0.45\textwidth]{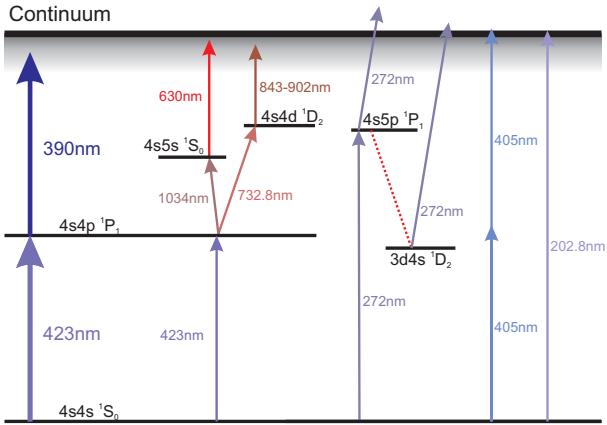}
\vspace{5mm}\caption{Selection of realized calcium photoionization
schemes: Resonance enhanced excitation of $^{40}$Ca to high-lying
Rydberg states via the 4s4p $^1$P$_1$-state \cite{gb01,wb97},
resonance enhanced continuum excitation via the 4s5p
$^1$P$_1$-state \cite{kd00} and non-resonant (two-photon)
excitation of high-lying Rydberg states \cite{aw77}.}
\label{fig:2stepscheme}
\end{center}
\end{figure}

In the typical experimental setup a narrow beam of neutral calcium
atoms from an oven is directed through the center of a Paul trap
where it is overlapped with the beams of the cooling lasers and
the photoionization light sources. The state of the art for
ionizing calcium atoms in ion trapping experiments is set by the
implementation of a two-photon resonance-enhanced photoionization
scheme with diode laser sources at 423nm and 390nm
\cite{gb01,ls04}. The idea of the scheme, as shown in Fig.
\ref{fig:2stepscheme}, is to excite the neutral calcium atoms in
the thermal beam resonantly on the 4s$^2$ $^1$S$_0\rightarrow$
4s4p $^1$P$_1$ transition and then further to high-lying
Rydberg-states which are subsequently Stark-ionized in the
electric field of the ion trap \cite{gb01,gk89}.

The S-P dipole transition has the broadest linewidth of all
transitions in Ca I (35.4MHz) and can easily be saturated with a
laser tuned to resonance \cite{ls04}. To excite the electron from
the $^1$P$_1$-state into the continuum above the first ionization
limit, a second photon of wavelength $\lambda\leq389.89$nm is
necessary, as e.g. realized in \cite{nb98}. However, high
ionization probability is also reached by exciting high-lying
Rydberg-states ($n\approx 40$) below the first ionization limit
($\lambda\geq389.89$nm), which are efficiently ionized by the
electric field of the Paul trap \cite{dk75}.

This technique allows for loading rates of more than 100 ions per
second \cite{gb01}. Nevertheless, for experiments with single ions
much lower rates are preferable, in order to easily control the
number of ions loaded into the trap. This is achieved with
low-power photoionization light sources and low-flux atomic beams.
Only for trapping of rare isotopes, slightly higher rates may be
necessary to beat charge exchange processes with $^{40}$Ca from
the atomic beam \cite{ls04}.

It was observed that even if the 390nm laser is operated below its
lasing threshold current, its weak incoherent emission was
sufficient to load ions \cite{ls04}. The efficiency of this
process was observed to be independent of the linewidth, as a
result of the broad autoionizing resonances in the vicinity of the
first ionization limit \cite{dh60,ch71,ns68}. Although the
excitation of Rydberg states is the bottleneck of the scheme,
sufficiently high loading rates to trap clouds of various calcium
isotopes, e.g. $^{43}$Ca$^+$, were reached in \cite{tu05,tu07}
using two high-power LEDs in a double-pass imaging configuration.

Despite the satisfactory performance of the two-photon scheme, its
implementation is hindered by the fact that 423nm and 390nm laser
diodes have become unavailable. The lasers used in current
experiments originate from early production runs of room
temperature current-injection III-IV nitride devices, where the
wavelength of the produced diodes varied unpredictably over a
range of 380-450nm \cite{n97}. Meanwhile the leading manufacturers
have refined their design and manufacturing techniques and
exclusively serve the consumer electronics and optical data
storage market with UV laser diodes complying with the Blu-ray
disc format (405nm). This situation demands an alternative
photoionization scheme for experiments with single ions, which
achieves high ionization efficiency and isotope selectivity with
reasonable effort and cost.

Previously, photoionization of calcium atoms has also been
demonstrated with excitation schemes, e.g. see Fig.
\ref{fig:2stepscheme}, which exhibit high efficiency and isotope
selectivity but rely on complex and expensive laser systems
\cite{kd00,md04,hgd07,dm88,dt04,aw77,bt83}. Apart from these
established schemes, an interesting possibility is using a
low-cost, free running 405nm laser diode for the excitation close
to the first ionization limit \cite{vb06}. For calcium this could
be achieved via a non-resonant two-photon transition or direct
excitation of high-lying Rydberg states after second harmonic
generation of $\lambda\approx202.5$nm light, e.g. in
KBe$_2$BO$_3$F$_2$ \cite{cw08}.

We pursued a more straightforward approach based on a combination
of a frequency-doubled infrared diode laser and a high power LED
for a novel implementation of the two-photon resonance-enhanced
scheme. To generate narrow-bandwidth 423nm radiation, we use a
commercially available infrared diode laser and second harmonic
generation (SHG) in a nonlinear crystal. This has been
demonstrated previously with bulk crystals but due to the
restrictions of birefringent phase matching in the UV, sufficient
power was only reached with more complex setups involving
resonance-enhancement cavities or power amplified master lasers
\cite{wb97,dt03,oh99}. Simpler setups for efficient second
harmonic generation can be realized with quasi-phase matched
materials, i.e. periodically poled crystals \cite{tgr03} or
waveguide structures \cite{pf02}. While waveguide structures have
higher conversion efficiencies, periodically poled crystals are
less expensive and easier to handle in terms of temperature
management and pump light coupling.

The experience with a sub-threshold 390nm-laser \cite{ls04} and
light emitting diodes \cite{tu05,tu07} also led us to implementing
the second excitation step from the $^1$P$_1$-state into the
continuum with a high-power LED and a custom designed optical
system. The two systems will be described in more detail in the
following sections.

\section{Second harmonic generation at 423nm in ppKTP}
\label{sec:423}

\begin{figure}[!t] \begin{center}
\includegraphics[width=0.45\textwidth]{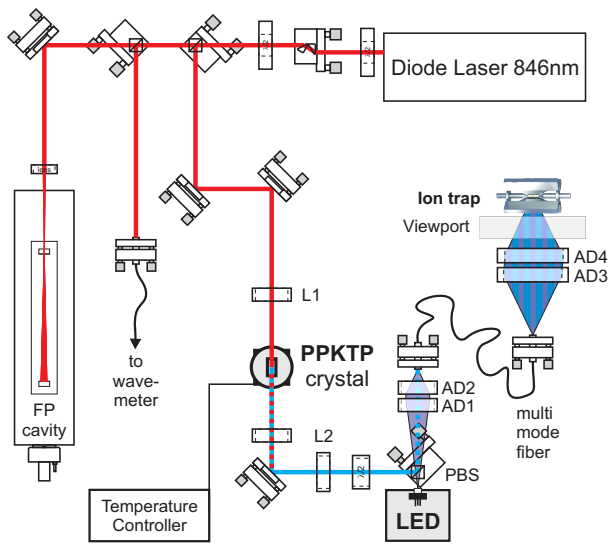}
\caption{Setup of photo-ionization light sources. The wavelength
and mode behavior of the 846nm pump laser are monitored with a
wavemeter and a Fabry-Perot (FP) cavity. The temperature of the
ppKTP crystal is set and stabilized with a commercial temperature
controller (PID-1500, Wavelength electronics). Lenses L1 and L2
are used to adjust the foci of the fundamental 846nm beam in the
crystal and the SHG 423nm beam in the multi-mode fiber,
respectively. Achromatic doublets AD1 \& AD2 and AD3 \& AD4 are
used to focus both 390nm and 423nm light into the multi-mode fiber
and from there to the trap center,
respectively.}\label{fig:pisetup}\end{center}
\end{figure}

For the generation of blue and ultraviolet radiation, periodically
poled potassium titanyl phosphate (ppKTP) has turned out to be
very attractive because stable output can be achieved even at
relatively high pump powers \cite{lp08}. Here we demonstrate how
to produce sufficient 423nm radiation to saturate the S-P
transition in Ca I by single pass frequency doubling of a
single-mode diode laser in a ppKTP crystal. The setup is displayed
in figure~\ref{fig:pisetup}.

As a pump source we use a grating-stabilized extended-cavity diode
laser (Toptica DL100) which has 3-4MHz linewidth. The SHG
radiation will have only half the linewidth of the pump beam
\cite{ha91} and is hence well suited for isotope-selective
photoionization. A drawback of using diode lasers for SHG is their
elliptical beam shape \cite{lp08}, which reduces the conversion
efficiency. To optimize the second harmonic power we therefore use
an anamorphic prism pair (see Fig. \ref{fig:pisetup}) to achieve a
circular collimated beam. We note that the beam shape is a
critical parameter for efficient second harmonic generation
because only fundamental modes for which the phasematching
conditions is fulfilled contribute to the power in the second
harmonic.

For a fundamental beam at $\lambda_{\omega}=846$nm we calculate
the poling period $\Lambda=4.05\mu$m in (biaxial) KTP from the
Sellmeier equations. Since the second harmonic power increases
linearly with crystal length we decided for a crystal of 20mm
length (crystal dimensions: $20\times5\times1$mm). Taking into
account irregularities of the poling period due to imperfect
control over the fabrication process, the beam shape of the
fundamental light and the high absorption coefficient at 423nm
($\sim10\%$cm$^{-1}$ \cite{tgr03}), longer crystals do not
necessarily lead to significantly higher single-pass SHG output
power.

The crystal \cite{ppktp} is held in a Brass holder, attached to a
kinematic prism mount which allows for tilting the crystal
entrance surface and hence the optical axis relative to the
incoming fundamental beam in horizontal and vertical direction. To
find the optimum trajectory of the focused pump beam through the
crystal the holder is mounted on a stack of translation stages
allowing us to precisely position the crystal in all three spatial
directions.

\begin{figure}[!t] \begin{center}
\includegraphics[width=0.45\textwidth]{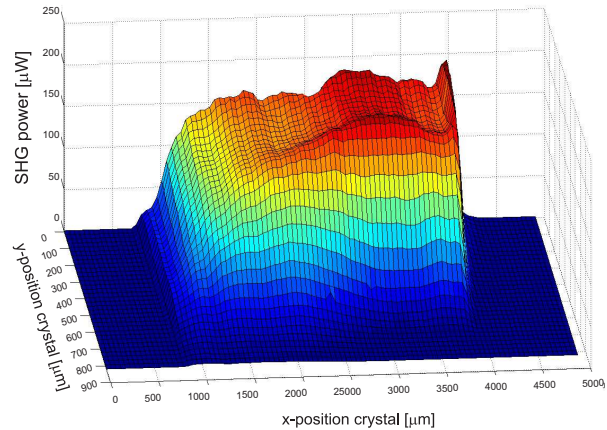}
\caption{Single pass SHG output power as a function of the entry
position of the pump beam into the ppKTP
crystal.}\label{fig:shgprofile}\end{center}
\end{figure}

To maximize the power of the second harmonic we set the focus
according to the Boyd-Kleinman theory \cite{bk68}. When scanning
the position of the pump beam focus by translating the crystal
along the direction of propagation we find a weak position
dependence of the second harmonic power which reaches its maximal
value at the center of the crystal, as expected \cite{bk68}. To
characterize the lateral position dependence we translate the
crystal perpendicular to the beam, i.e. horizontally (referred to
as x in Fig.\ref{fig:shgprofile}) and vertically (y). As shown in
figure \ref{fig:shgprofile}, we find that the achievable second
harmonic output power depends on the position where the pump beam
enters the crystal and is highest close to the top surface. In
x-direction, the poling was only applied to the central 3mm of the
crystal and the edges on either side remained unpoled. The highest
SHG conversion efficiency is achieved in the $2500\times400\mu$m
region below the top surface, Fig. \ref{fig:shgprofile}, and shows
variations of approximately $40\%$.

These findings are explained by the electric field poling process
\cite{cl03}. For our crystal the periodic electrode pattern was
deposited on the top surface from where the domains grew in the
shape of a needle towards the other electrode on the bottom side.
Hence, close to the top surface the domains are best defined and
highest uniformity of the quasi-phasematching grating is achieved,
which makes second harmonic generation most efficient in this
region. Deeper inside the crystal the domains spread out, split
and may not reach the bottom side. However, it is not
straightforward to draw more detailed conclusions about the
quality of the periodically poled medium from these measurements
because the spectrum of the pump is too narrow \cite{ha91}.

Importantly, efficient second harmonic generation is reached over
a large crystal entry surface area (Fig. \ref{fig:shgprofile})
such that the setup is relatively insensitive to alignment and
requires low maintenance in practice.

Since the indices of refraction depend on temperature, one has to
be able to precisely control the crystal temperature to fulfill
the phasematching conditions for the desired fundamental
wavelength. We achieve this with a thermoelectric element glued to
the crystal holder. To set and stabilize the crystal temperature
with high accuracy we use a commercial PID circuit controlling the
current through the thermoelectric element. This temperature
controller provides low-noise current and keeps the crystal
temperature stable within 5mK over 24 hours. As sensor input we
use a negative temperature coefficient thermistor (NTC) glued to
the crystal holder.

\begin{figure}[!b] \begin{center}
\includegraphics[width=0.45\textwidth]{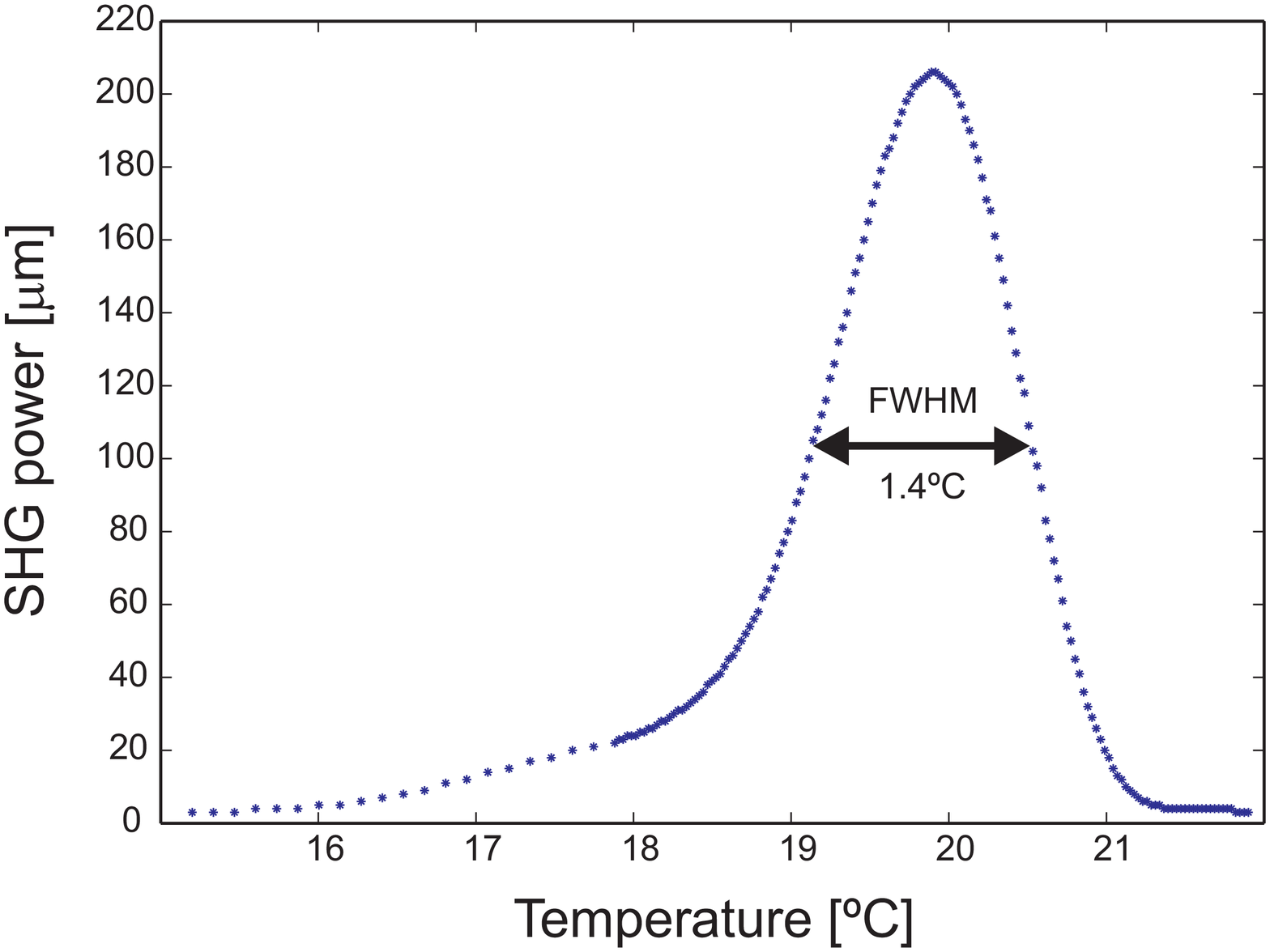}
\caption{Temperature dependence of second harmonic generation in
ppKTP-crystal at $846$nm. The crystal temperature is controlled
with a PID-regulated thermoelectric and varied in steps of
28mK.}\label{fig:shgvst}\end{center}
\end{figure}

To determine the optimal temperature for second harmonic
generation to 423nm, we vary the crystal temperature while
monitoring the frequency-doubled output power after a color-filter
(BG39, 2mm) eliminating the residual pump light. The maximum power
was reached at $19.9^{\circ}$C, see Fig.\ref{fig:shgvst}, a
convenient working point above the dew point but easily reachable
without the need for an oven as in other implementations
\cite{tgr03}. The obtained full width at half maximum is
$1.4^{\circ}$C, which shows that our simple temperature control
system guarantees sufficient stability for reliable operation with
no need for periodic readjustment.

The observed asymmetry in the temperature dependence is due to a
spread of the wavevectors in the focused beam, since the
phasematching conditions are fulfilled at different temperatures
for different wavevector components of the pump beam \cite{gl98}.
For a beam which is not as tightly focused, the phasematching
conditions should be approximately constant over the entire
crystal length and one expects a sinc$^2$ behavior \cite{kl97}.

The wavelength of the SHG output is set to resonance with the
S-P-transition by tuning the fundamental wavelength. As shown in
Fig.\ref{fig:pisetup} we split off a small fraction of the pump
laser to monitor its wavelength by a wavemeter. By setting the
angle of the feedback grating of the extended cavity diode laser
we vary the pump wavelength until we observe fluorescence on an
EMCCD camera from an intense atomic calcium beam through the trap,
see Fig. \ref{fig:ions} a). This step was performed only once, and
the wavemeter reading at maximal fluorescence from the atomic beam
is used henceforth as a reference. Before each loading sequence, a
slight (piezo-controlled) readjustment of the fundamental
wavelength is then sufficient to guarantee efficient excitation of
the $^{40}$Ca S-P transition. For loading single ions on a daily
basis we work with much weaker atomic beams, i.e. lower oven
currents, where the atomic fluorescence is hardly visible on the
camera.

The Fabry-Perot cavity shown in Fig.\ref{fig:pisetup} is used to
monitor the (single) mode behavior of the pump laser but could in
principle also be used to integrate the pump laser into a
transfer-lock frequency-stabilization scheme used for all other
laser sources in our experiment \cite{Felix}. This may for example
be desirable in experiments where higher isotope selectivity is
required. However, it turned out to be unnecessary to do so for
experiments with $^{40}$Ca. Due to the high intrinsic stability of
our external cavity diode laser the drift during loading is
relatively small compared to the transition linewidth (35.4MHz)
and the isotope shifts ($>400$MHz) \cite{ls04}.

We generate $315.5\mu$W of narrow-bandwidth 423nm radiation from a
119mW pump beam when all parameters are optimized. This
corresponds to a single pass conversion efficiency of $\eta_{\rm
eff}=1.11\%$(W cm$)^{-1}$. Taking into account transmission loss
in the optical components, the intensity of the 423nm light
focused to a spot size of $250\mu$m at the trap center reaches
approximately $I_{\rm 423nm}=$5mW/mm$^2$, which is sufficient to
saturate the S-P-transition of Ca~I atoms ($I_{\rm
sat}=3.7$mW/mm$^2$ \cite{ls04}). Note, that we use a
single-frequency laser as compared to previous experiments with,
e.g., Ti:sapphire lasers \cite{tgr03}, where a large number of
modes enhances the conversion efficiency by a factor of two
\cite{ha91}. Taking further into account the spatial intensity
distribution of the pump beam, our SHG conversion efficiency is
comparable to results obtained with significantly more complex
setups. In \cite{lp08} a power amplified single-frequency pump
laser and crystals of different length gave similar conversion
efficiency values at a different wavelength and slightly higher
efficiencies are reported for 423nm SHG in \cite{tgr03}, where a
larger pump linewidth and a better beam profile were used.

\section{Photoionization with a high-power LED}
\label{sec:led}

Light emitting diodes have shown to be suitable to load ion clouds
of different isotopes \cite{tu05,tu07} and single $^{40}$Ca$^+$
ions \cite{ls04}. For experiments where low loading rates are
acceptable, a broad-bandwidth radiation source of relatively low
power (i.e. far below saturation of the involved transitions) is
sufficient, since many transitions of partially overlapping
linewidths and broad autoionizing resonances \cite{dt04,bb04} are
available for the (electric field assisted) excitation into the
continuum. We chose to use a high-power UV-LED from Nichia
(NCCU001) to achieve loading of single ions, and we designed an
optical system tailored to the LED emission characteristics. The
NCCU001 has significantly higher optical output power than
previous UV-LED models, specified between 70 and 85mW \cite{led}
and does not require a current controller.

The suitability of light emitting diodes for photoionization
studies in the past was limited \cite{ls04} due to their very
broad emission spectra, wide emission angles and often large light
emitting areas, characteristics which also the NCCU001 displays.
Its emission spectrum, centered around 380nm, has a width of
$\Delta\lambda_{\rm FWHM}=30$nm, of which only a small fraction
excites transitions from the $^1$P$_1$-state to high-lying Rydberg
states just below the first ionization limit. However, the
efficiency of such excitation depends only weakly on linewidth in
this wavelength region. The light emitting area is $1\times1$mm
large and its angular emission pattern extends over a solid angle
of almost $2\pi$. Hence, the main challenge in using an LED
instead of a laser diode for photoionization of Ca is to collect
and focus its optical output such that sufficiently high
intensities are reached within an appropriate volume at the trap
center.

For efficient photoionization the atomic beam, the 423nm laser and
the LED radiation have to overlap in the same volume. We thus
consider a spot size of $200\mu$m at the center of the trap as
suitable. This is sufficiently small to reach the saturation
intensity on the atomic S-P transition with the available 423nm
power, avoids illumination of the trap electrodes, while covering
a large enough part of the atomic calcium beam for the excitation
process to be efficient.

To reach high intensity of the LED radiation at the trap center,
the optical system has to be optimized for efficient collection
from the light emitting surface and for focussing to the desired
spot size with low aberrations. The collimation of light from a
highly divergent source requires the use of lenses of high
numerical aperture. To image the whole LED surface of object size
1mm to the envisioned spot size would require an optical system of
(de)magnification $m=1/5$. This implies that the object distance
(LED to optical system) has to be five times larger than the image
distance (optical system to demagnified image) and accordingly the
numerical aperture (NA) that can be collected from the LED is
limited to a fifth of that of the imaging lens. The main
restriction on the design is then given by the imaging numerical
aperture, which is limited by the 43mm distance between the vacuum
window and the trap center. Also the UV-transmission properties
and aberrations of real lenses impose further restrictions, since
high-NA lenses show high spherical and chromatic aberrations.

We performed simulations with an optical system design package
(ZEMAX Development Corp.) to find a suitable imaging system based
on commercially available, achromatic doublet lenses. Depending on
the particular design and the chosen materials available from
different manufacturers, aberrations were found to vary
significantly but generally become very large for optical systems
with numerical apertures NA$\geq0.25$ and focal lengths
$f\geq40mm$. We use one inch diameter achromatic doublets (denoted
as AD1 and AD2 in Fig. \ref{fig:pisetup}) of $f_{\rm AD1}=200$mm
and $f_{\rm AD2}=40$mm focal lengths which collect the light
emitted into a solid angle of $\Omega\approx 0.1\%\times 2\pi$ and
focus it to a spot size of $215\mu$m, corresponding to a
magnification $m\approx1/5$. By integrating the LED's angular
emission pattern over the solid angle covered by the imaging
system and assuming a total horizontally polarized optical output
power of 40mW, we estimate the collected LED power in the whole
emitted wavelength range as $P_c=210\mu$W.

It is useful to couple both light sources into a multi-mode fiber
which is conveniently delivered to the trap, guarantees good
spatial overlap of the LED and 423nm beams and occupies minimal
optical access to the trap center. We use a multi-mode fiber of
$200\mu$m core diameter and numerical aperture NA$=0.22$, which
adapts well to the above imaging parameters. To overlap the 423nm
laser beam and the LED radiation we use a polarizing beam
splitter. With lens L2, shown in the setup of
Fig.~\ref{fig:pisetup}, the coupling of both beams to the
multi-mode fiber is optimized independently. The one-to-one
imaging of the fiber core to the trap center is then realized with
achromatic doublets (denoted as AD3 and AD4 in Fig.
\ref{fig:pisetup}) of 75mm focal lengths and 50mm diameter. Due to
aberrations the actual spot size at the trap center is almost
$250\mu$m. The optimal distances between the optical components to
simultaneously image the 423nm laser and the LED's light emitting
surface to the fiber input and the fiber output to the trap center
is determined from the ZEMAX-simulations. The output fiber coupler
and the imaging lenses are mounted on translation stages to
overlap the focal spot of the photoionization beams with the
atomic beam at the trap center.

To estimate the LED power available for photoionization at the
trap center we consider the wavelength range $\lambda=365-391$nm
of the total spectral power distribution, i.e. assuming that
Rydberg states down to $n=40$ ($\lambda(n=40)\approx391$nm) can be
ionized in the electric field of the Paul trap. Taking further
into account 75\% transmission through the 2m long multi-mode
fiber, the two achromatic doublet lenses and the 10mm thick
viewport we obtain $P_{365-391\textmd{nm}}\approx150\mu$W, i.e. an
intensity of $I_{365-391\textmd{nm}}\approx3$mW/mm$^2$, of which
$2.3\mu$W are emitted below the first ionization threshold. A more
precise treatment would have to take into account the complex
level structure of atoms close to the ionization limit,
autoionizing resonances \cite{dt04,bb04}, series perturbations and
core polarization effects \cite{hv99}.

\section{Loading clouds and single ions}
\label{loading}

\begin{figure}[!t] \begin{center}
\includegraphics[width=0.45\textwidth]{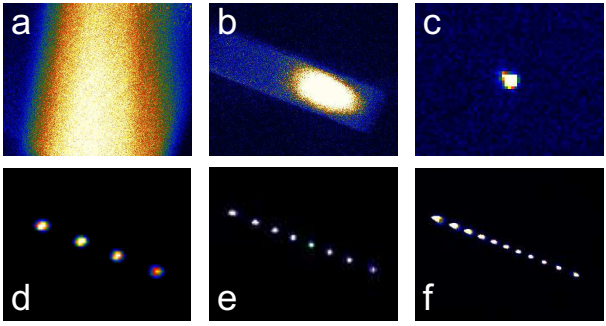}
\vspace{5mm}\caption{a) Fluorescence from atomic calcium beam, b)
trapped $^{40}$Ca$^+$ cloud seen between the trap electrodes, c)
single $^{40}$Ca$^+$, d)-f) strings of
$^{40}$Ca$^+$.}\label{fig:ions}\end{center}
\end{figure}

In our setup the atomic calcium beam is produced from a
resistively heated stainless steel tube (2.5mm diameter) filled
with granular calcium. To collimate the beam evaporating from this
oven we use an orifice of 1mm diameter at a distance of 19.8mm
from the trap center. For trapping single ions we usually work
with oven currents of 4.5A to 5A. The cooling lasers, fluorescence
collection and detection are described elsewhere
\cite{Felix,ge09}. With the photoionization light sources and the
cooling lasers aligned to overlap with the calcium beam at the
center of the linear Paul trap, the loading of ions is monitored
in real time on an EMCCD camera. We are able to load single
$^{40}$Ca$^+$ ions, crystallized ion strings and large ion clouds,
see Fig.~\ref{fig:ions} b)-f).

The linewidth and intensity of the focused 423nm radiation are
adequate for isotope selective photoionization loading and
isotopically pure strings of $^{40}$Ca$^+$ are routinely loaded.
The only parameter that has to be adjusted for each loading
sequence is the wavelength of the pump laser due to long term
drifts. Using the wavemeter, the setup of the photoionization
light sources requires no maintenance on a day to day basis.

While it was not necessary for our application, higher frequency
control for isotope-selective S-P excitation may be reached by
integrating the monitoring cavity of the 846nm pump laser (see
Fig. \ref{fig:pisetup}) into the transfer locking scheme described
in \cite{Felix}. However, different isotopes may also require
other laser cooling schemes \cite{ls04}.

The setup we used for photoionization was built with commercially
available light sources and off-the-shelf parts and is
maintenance-free. The optical system is tailored to the LED
emission characteristics and only occupies optical access to the
trapping region along one direction. Optimal overlap of the LED
and the 423nm SHG laser light at the trap center is achieved by
coupling both sources into a multi-mode optical fiber. Our
implementation is significantly simpler than those of previous
photoionization studies for Ca$^+$ trapping experiments, which
used a 423nm diode laser, two high-power UV-LEDs and an imaging
system occupying optical access along five directions to trap ion
clouds \cite{tu05,tu07}.

\section{Observation of Quantum Jumps}

\begin{figure}[!t] \begin{center}
\includegraphics[width=0.4\textwidth]{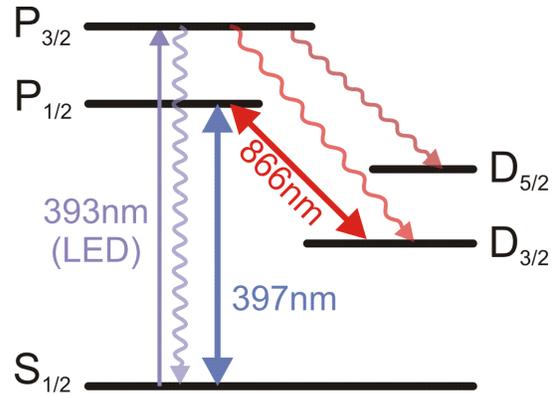}
\vspace{5mm}\caption{Level scheme of
$^{40}$Ca$^+$.}\label{fig:ca+}\end{center}
\end{figure}

We note that the LED spectrum is broad enough to not only excite
Ca I atoms from the $^1$P$_1$-state to states in the vicinity of
the first ionization limit ($\lambda\approx 389.89$nm), but it
also contains spectral components which are resonant with the
S$_{1/2}$-P$_{3/2}$ transition ($\lambda= 393$nm) in
$^{40}$Ca$^+$, see figure \ref{fig:ca+}. We observe that the LED
power emitted in this frequency range is sufficient to
occasionally change the internal quantum state of a laser cooled
calcium ion.

Once a single ion or a string of ions is loaded into the trap and
the 397nm cooling- and 866nm repump-lasers are tuned close to
resonance with the S$_{1/2}$-P$_{1/2}$ and the P$_{1/2}$-D$_{3/2}$
transitions (see Fig. \ref{fig:ca+}), continuous photon scattering
is observed. We then switch on the LED again and observe that the
resonance fluorescence abruptly ceases to the dark count level and
sets in again in irregular intervals on a time scale of around 1s.
Such random fluorescence extinction occurs when the ion is
transferred into a metastable (shelving) state. Similar quantum
jumps were first observed in \cite{nd86,st86,bw86} and demonstrate
the quantum nature of resonance fluorescence as they allow for
directly monitoring the quantum state of an ion. Thus, the abrupt
changes in the fluorescence signal due to jumps to and from a
metastable state have practical applications, e.g. for highly
efficient detection of atomic transitions or states (electron
shelving).

In our case an ion absorbs some of the photons emitted by the LED
which are resonant with the S$_{1/2}$-P$_{3/2}$ transition, see
figure \ref{fig:ca+}. Once it is excited to the P$_{3/2}$-state we
distinguish two decay channels. With high probability it will
spontaneously decay to the S$_{1/2}$- or D$_{3/2}$-states from
where it reenters the cooling cycle. Since the P-states have very
short lifetimes ($\tau\approx 7ns$), this detour from the cooling
cycle via the P$_{3/2}$-state is not observed in the fluorescence
signal. However, spontaneous decay from the P$_{3/2}$-state may
also occur, with branching ratio $\Gamma_{\rm D_{5/2}}/\Gamma_{\rm
tot}=1/17$, to the metastable D$_{5/2}$-state, which is not part
of the cooling cycle. Unlike the P-states, the D$_{5/2}$-state has
a long lifetime of roughly one second and the cessation of the
fluorescence signal is observable with a CCD-camera or a
photodetector, before spontaneous decay from the D$_{5/2}$-state
takes the ion back to the S$_{1/2}$-ground state. From there it
reenters the cooling cycle and the fluorescence signal returns to
its previous value. The frequency of quantum jumps can be used to
optimize the alignment of the photoionization beams to the trap
minimum.

We also use the average rate $R_{\rm QJ}\approx0.5$Hz at which we
observe quantum jumps on the EMCCD camera to estimate the LED
power at the ion in the (non-Gaussian) focal spot area, $A=\pi
\omega^2_f$. For an unpolarized photon resonant with the ionic
S$_{1/2}$-P$_{3/2}$ transition the scattering probability is
$p_{\rm
sc}=f_{\textmd{S}_{1/2}\rightarrow\textmd{P}_{3/2}}\cdot\sigma/A$,
where $\sigma=\lambda^2/2\pi$ is the on-resonance cross section
and $f_{\textmd{S}_{1/2}\rightarrow \textmd{P}_{3/2}}=0.626$ the
oscillator strength of the S$_{1/2}\rightarrow$P$_{3/2}$
transition. From the flux of resonant photons and the photon
energy we find the power
\begin{equation*}
P'_{SP}=\frac{R_{\rm QJ}\Gamma_{\rm tot}}{p_{\rm
sc}\Gamma_{\textmd{D}_{5/2}}}\cdot\frac{hc}{\lambda}=14{\rm pW},
\end{equation*}
A trapped ion thus experiences an intensity of $I_{\rm
SP}'=280$pW/mm$^2$ on the S$_{1/2}\rightarrow{\rm P}_{3/2}$
transition from the LED light field. Using this value to integrate
the spectral power distribution numerically over the wavelength
range $\lambda=365-391$nm yields the power available for
photoionization at the trap center
$P_{365-391\textmd{nm}}'=144\mu$W, which is in very good agreement
with the independently obtained value in section \ref{sec:led}.

\section{Conclusions}

We have implemented a two-photon resonance-enhanced
photoionization scheme to trap clouds, strings and single
$^{40}$Ca$^+$ ions. We achieve efficient photoionization of
calcium atoms in a thermal beam with a high power ultraviolet
light emitting diode and single-pass frequency doubling of an
infrared extended cavity diode laser in periodically poled KTP,
i.e. employing commercially available, inexpensive and robust
light sources. With a suitably designed optical system, tailored
to the LED emission characteristics and based on off-the-shelf
components, we achieve intensities of approximately $I_{\rm
423nm}=5$mW/mm$^2$ and $I_{\rm 365-391nm} = 3$mW/mm$^2$ over a
spot size of $250\mu$m at the trap center. The setup is simple,
maintenance-free, occupies minimal optical access to the trapping
region and can easily be adapted to trapping of rare calcium
isotopes.

\begin{acknowledgement}

We acknowledge support by the European Commission (SCALA, Contract
No.\ 015714) and by the Spanish Ministerio de Educaci\'on y
Ciencia (QOIT, Consolider-Ingenio 2010 CSD2006-00019; QLIQS,
FIS2005-08257; QNLP, FIS2007-66944). C.S. acknowledges support by
the Commission for Universities and Research of the Department of
Innovation, Universities and Enterprises of the Catalonian
Government and the European Social Fund.

\end{acknowledgement}

\bibliographystyle{unsrt}
\bibliography{bibliography-pi}

%
%
%

\end{document}